\documentclass[12pt,a4paper,final]{iopart}

\usepackage{iopams}
\usepackage{graphicx}
\usepackage[breaklinks=true,colorlinks=true,linkcolor=blue,urlcolor=blue,citecolor=blue]{hyperref}

\expandafter\let\csname equation*\endcsname\relax
\expandafter\let\csname endequation*\endcsname\relax
\usepackage{amsmath,amsfonts}
\usepackage{tikz}
\usepackage{color}
\usetikzlibrary{arrows, automata}
\usepackage[normalem]{ulem}

\begin{document}

\title[Adaptive Strategies to Fast Multipole Method in Photoionization Calculation]{Adaptive Strategies to Fast Multipole Method in Photoionization Calculations}

\author{Bo Lin}
\address{Department of Mathematics, National University of Singapore, 10 Lower Kent Ridge Road, Singapore 119076}
\ead{linbo@u.nus.edu}

\author{Chijie Zhuang}
\address{State Key Lab of Power Systems and Department of Electrical Engineering, Tsinghua University, Beijing 100084, China}
\ead{chijie@tsinghua.edu.cn}

\begin{abstract}
Recently, a new framework to compute the photoionization rate in streamer discharges accurately and efficiently using the integral form and the fast multipole method (FMM) was presented. This paper further improves the efficiency of this framework with adaptive strategies. The adaptive strategies are based on the magnitude of radiation and the electric field during the streamer propagation, and are applied to the selection of the source and target points. The accuracy and efficiency of this adaptive FMM are studied quantitatively for different domain sizes, pressures and adaptive criteria, in comparison with some existing efficient approaches to compute the photoionization. It is shown that appropriate adaptive strategies reduce the calculation time of the FMM greatly, and maintain the high accuracy that the numerical error is still much smaller than other models based on partial differential equations. The performance of the proposed adaptive method is also studied for a three-dimensional positive streamer interacting problem with a plasma cloud.

\end{abstract}

\vspace{2pc}


\section{Introduction}
Streamer discharge is an important building block of gas discharges. As a natural phenomenon, streamer discharge can be observed and captured directly in high altitude as sprite discharges \cite{liu2015sprite, sprite2020}. It is also the precursor of sparks \cite{spark2017, spark2018} and lightning leaders \cite{stoleader2013, aplLeader2017, PRL2020Leader} in lower altitude. As a product of streamer discharge, the non-thermal atmospheric plasma paves the way to a large number of chemical reactions in applications such as pollution control \cite{yan2001chemical, ChemPhys2008}, solubilization of carbon nanomaterial \cite{2006Carbon, 2007Carbon} and disinfection \cite{disinfection2018, foods2019}. More background information of streamer discharge could be found in a recent review in \cite{UteReview2020}.

Free electrons and high electric field are two important components in streamer discharge. During the propagation of streamer, the electric field are enhanced locally ahead of it. Under this enhanced field, free electrons are accelerated to gain sufficient energy and impact with the gas molecules to generate new electrons and ions, which elongate the streamer channel.

When streamer moves against the direction of electron drift, which is called positive streamer or cathode-directed streamer, it requires external free electrons in front of its head. In air or other oxygen-nitrogen mixtures, the free electrons are provided by photoionization \cite{PhysRevMeek1940, Kulikovsky1997}, which makes the photoionization significant in the propagation of positive streamers. Moreover, the photoionization has an impact on streamers in their velocities \cite{PhotoVel2018, PhoVel2021} and branching behaviors \cite{xiong2014branching, bagheri2019effect, marskar20203d}.

The importance of photoionization has attracted continuous efforts for its modelling and simulation. Based on the experimental data and understanding of previous researchers, Zheleznyak \textit{et al.} derived the classical integral model for oxygen-nitrogen mixture in \cite{zhelezniak1982} with detailed coefficients, which is widely used in the simulation of streamer discharges \cite{PhoVel2021, stephens2018practical, Analytic1d2020}. This integral model is improved and extended in \cite{pancheshnyi2014, framework2019}. However, its direct implementation requires a large amount of computation especially in three dimensions (3D), since the calculation of the photoionization at one specific point requires a quadrature over the whole simulation domain. 

To facilitate the direct calculation in 3D, there are typically two approaches. The first approach \cite{segur2006, luque2007, bourdon2007} is to expand the kernel function in the integral and convert the integral into a small number of partial difference equations (PDEs). Together with some efficient PDE solvers, this approach reduces computation time greatly \cite{tmag2020}. Unfortunately, as stated in \cite{segur2006, bourdon2007}, the accuracy of this approach highly depends on the fitting coefficients in the expansion, the absorption coefficients in the photoionization, the size of domain and the boundary conditions of PDEs.

The second approach is to reduce the quadrature points by adopting a coarser mesh to compute the photoionization \cite{PhoVel2021, bourdon2007, Hallac2003} or confining the integral into a subdomain around the head of streamer \cite{pancheshnyi2001role}. This approach provides a straightforward way to reduce the computation time, and it keeps the integral form which can be implemented in non-constant gas \cite{PhoVel2021}. However, its efficiency and accuracy depend on the choice of coarse grid and subdomain, therefore its error is difficult to estimate \cite{segur2006}.

Instead of following the two approaches, we facilitate the direct calculation by means of the fast multipole method (FMM, \cite{ying2004}) recently in \cite{Lin2020psst}. This new approach is based on the quadrature of the classical integral model, which is free of fitting coefficients and additional boundary conditions. Moreover, it is easy to extend to other integral forms such as the integral framework in \cite{framework2019}. The accuracy and computation time of this approach are compared with the PDE-based methods in \cite{Lin2020psst}, which shows the new approach is much more accurate than the PDE-based methods under similar computational cost. Furthermore, it exhibits more robustness with respect to domain sizes and pressures.

Although the computation time of our new approach is comparable to the PDE-based approach, the time usage is mildly higher, especially comparing to the Helmholtz approximation in \cite{luque2007} with zero boundaries (using three-term expansion). On the other hand, the high accuracy of the new approach might not be significant when simulating the atmospheric streamer discharge in strong electric fields, since the photoionization has weaker influence compared to the the impact ionization. This motivates us to accelerate our new approach with acceptable sacrifice on its accuracy.

Inspired by the idea of reducing quadrature points in the second approach, we improve the efficiency of our FMM approach with adaptive strategies in this paper. Three adaptive strategies are adopted to select only a portion of quadrature points which have more contribution to the photoionization. Instead of choosing a coarser grid or a confined box, our strategies are based on the values of source radiation and electric field during streamer propagation. These strategies are: (i) choosing source points with higher radiation; (ii) choosing target points according to the magnitude of electric field; (iii) skipping some steps in the FMM algorithm when the relative value of source radiation to the target photoionization is small. It is easy to implement these strategies, and numerical experiments illustrate that they make the FMM more efficient by little sacrifice on the accuracy.

The rest of this paper is outlined as follows. Section \ref{model} reviews the theories and formula of the classical integral model, the PDE-based approximations and the fast multipole method. Then, the adaptive strategies for the fast multipole method are introduced in section \ref{fmm}. The performance of the fast multipole method with adaptive strategies and its quantified comparison with other methods are conducted in section \ref{comparison} for computing photoionization; and in section \ref{comparison1} for simulating streamer discharges with localized plasma cloud. Finally, conclusions and future works are drawn in section \ref{conclusion}.

\section{Model formulation}\label{model}
Some deterministic approaches for photoionization calculations and the fast multipole method were reviewed in \cite{Lin2020psst}. To make the contents self-contained in this paper, we review these approaches again but in a more concise way.

\subsection{Classical integral photoionization model by Zheleznyak \textit{et al.}}\label{classint}

A classical model of the photoionization is derived by Zheleznyak \textit{et al.} \cite{zhelezniak1982}, which describes the photoionization rate at position $\vec{x}= (x, y, z)^T$ in a collector chamber $V$ by
\begin{equation}
S_{\rm ph}(\vec{x}) = \iiint_{V'} \frac{I(\vec{y})g(|\vec{x}-\vec{y}|)}{4\pi |\vec{x}-\vec{y}|^2}\mathrm{d}\vec{y},
\label{integral}
\end{equation}
where $V'$ is the source chamber.
$I(\vec{y})$ is proportional to the intensity of radiation:
\begin{equation}
I(\vec{y}) = \xi \frac{p_q}{p+p_q} \frac{\omega}{\alpha} S_i(\vec{y}),
\label{Ifun}
\end{equation}
where $\xi$ is the photoionization efficiency, $p$ is the gas pressure, $p_q$ is the quenching pressure, $\omega$ is the excitation coefficient of emitting states without quenching processes, $\alpha$ is the Townsend ionization coefficient, and $S_i$ is the ionization rate. The kernel function $g(r)=g(|\vec{x}-\vec{y}|)$ in (\ref{integral}) is given by
\begin{equation}
\frac{g(r)}{p_{_{O_2}}} = \frac{\exp(-\chi_{\min}\, p_{_{O_2}}r) - \exp(-\chi_{\max}\,p_{_{O_2}}r)}{p_{_{O_2}}r \ln(\chi_{\max}/\chi_{\min})},
\label{gfun}
\end{equation}
where $r=|\vec{x}-\vec{y}|$, $p_{_{O_2}}$ is the partial pressure of oxygen, $\chi_{\min}=0.035$\,cm$^{-1}$\,Torr$^{-1}$ and $\chi_{\max}=2$\,cm$^{-1}$\,Torr$^{-1}$ are the minimum and maximum absorption coefficients of O$_2$ in wavelength 980-1025\,\AA \cite{zhelezniak1982}, respectively. 

\subsection{Exponential or Helmholtz PDE approximation}\label{Helmholtz}
To avoid the direct computation of integral \eqref{integral}, the kernel function was approximated by the sum of the solutions of Helmholtz equations in \cite{luque2007,bourdon2007} as follows:
\begin{equation}
\frac{g(r)}{p_{_{O_2}}} \approx p_{_{O_2}} r \sum_{j=1}^{N_{E}} C_j \exp(-\lambda_j p_{_{O_2}} r),
\label{gexp}
\end{equation}
where $\lambda_j$ and $C_j$ ($1 \leq j \leq N_E$) are constants to be fit. With the approximation of kernel function, the photoionization rate in \eqref{integral} could be written as  
\begin{equation}
S_{\rm ph}(\vec{x}) \approx \sum_{j=1}^{N_{E}}C_j S_{{\rm ph},j}(\vec{x}),
\label{sphHel}
\end{equation}
where $S_{{\rm ph},j}(\vec{x})$ is the solution of the following modified Helmholtz equation
\begin{equation}
(-\Delta + (\lambda_j p_{_{O_2}})^2 ) S_{{\rm ph},j}(\vec{x}) = (p_{_{O_2}})^2I(\vec{x}).
\label{Hel}
\end{equation}
As a result, it suffices to solve $N_E$ modified Helmholtz equations numerically (e.g., using the multigrid-preconditioned FGMRES method \cite{lin2018}), which is more efficient compared to the direct quadrature of \eqref{integral}.

In this paper, we simply follow \cite{bourdon2007} to take $N_E=3$ in \eqref{gexp} and the associated coefficients $C_j$ and $\lambda_j$ therein. 
Zero boundary conditions are adopted for all $S_{{\rm ph},j}$ in \eqref{Hel}, as used in \cite{luque2007}.


\subsection{Three-group radiative transfer approximation}\label{Threegroup}
Radiative transfer equation and its approximations give another way to compute the photoionization rate using differential equations. In \cite{segur2006, bourdon2007,fvmrte2008}, the multi-group approximation to the radiative transfer equation is adopted, where the intensity of radiation $\Psi_j$ for the $j$-th group of spectral frequency satisfies
\begin{equation}
\vec{\omega} \cdot \nabla \Psi_j(\vec{x}, \vec{\omega}) + \kappa_j \Psi_j(\vec{x}, \vec{\omega}) = \frac{I(\vec{x})}{4 \pi c\, \xi}, \quad j = 1,\cdots,N_{\nu},
\label{rte}
\end{equation}
where $\vec{\omega} \in S^2$ is the solid angle, $\kappa_j = \lambda_j\, p_{_{O_2}}$ is the absorption coefficient in air and $c$ is the speed of light. Then the photoionization rate equals to the linear combination of 
the integral of $\Psi_j$ as
\begin{equation}
\begin{split}
S_{\rm ph}(\vec{x}) &= \sum_{j=1}^{N_{\nu}} A_j \,\xi\, p_{_{O_2}}c \int_{S^2} \Psi_j(\vec{x},\vec{\omega})
\mathrm{d}\vec{\omega}, \\
&=\sum_{j=1}^{N_{\nu}} A_j\, p_{_{O_2}} \iiint _{V} \frac{I(\vec{y}) \exp(-\lambda_j p_{_{O_2}}|\vec{x}-\vec{y}|)}{4 \pi |\vec{x}-\vec{y}|^2} \mathrm{d}\vec{y}.
\end{split}
\label{rteSph}
\end{equation}
Noting that \eqref{rteSph} is identical to (\ref{integral}) if
\begin{equation}
\sum_{j=1}^{N_{\nu}} A_j p_{_{O_2}} \exp(-\lambda_j p_{_{O_2}} r) = g(r),
\label{rtetoint}
\end{equation}
where $r=|\vec{x}-\vec{y}|$. The coefficients $A_j$ and $\lambda_j$ ($1 \leq j \leq N_{\nu}$) in \eqref{rtetoint} can be determined by fitting with $g(r)$ in \eqref{gfun}. In this paper, we adopt the coefficients from \cite{bourdon2007} for three-group ($N_{\nu}=3$) approximation.


It is more efficient to solve the differential equation \eqref{rte} instead of calculating the quadrature in \eqref{rteSph}. Besides directly solving the radiative transfer equation \cite{fvmrte2008}, the improved Eddington or SP$_3$ approximation \cite{larsen2002} is adopted in \cite{segur2006, bourdon2007, tmag2020} to approximate the isotropic solution by
\begin{equation}
\int_{S^2} \Psi_j(\vec{x},\vec{\omega}) \mathrm{d}\vec{\omega} = \frac{\gamma_2 \phi_{j,1} (\vec{x}) - \gamma_1 \phi_{j,2} (\vec{x})}{\gamma_2-\gamma_1},
\label{sp3sol}
\end{equation}
where $\gamma_n = \frac{5}{7}\left[ 1 + (-1)^n 3 \sqrt{\frac{6}{5}}\right]$ ($n=1,2$), and $\phi_{j,1} (\vec{x})$ and $\phi_{j,2} (\vec{x})$ satisfy the following two modified Helmholtz equations
\begin{align}
\left(- \Delta + \frac{(\lambda_j p_{_{O_2}})^2}{\mu_1^2} \right) \phi_{j,1}(\vec{x}) = \frac{\lambda_j p_{_{O_2}}}{\mu_1^2} \frac{I(\vec{x})}{c\,\xi}, \label{phi1} \\
\left(- \Delta + \frac{(\lambda_j p_{_{O_2}})^2}{\mu_2^2} \right) \phi_{j,2}(\vec{x}) = \frac{\lambda_j p_{_{O_2}}}{\mu_2^2} \frac{I(\vec{x})}{c\,\xi} , \label{phi2}
\end{align}
with the coefficients $\mu_{n}=\sqrt{\frac{3}{7} + (-1)^n \frac{2}{7}\sqrt{\frac{6}{5}}}$ ($n=1,2$). One efficient boundary condition for \eqref{phi1}-\eqref{phi2} is proposed in \cite{larsen2002, liu2007apl} as
\begin{align}
\nabla \phi_{j,1} \cdot \vec{n} + \alpha_1 (\lambda_j p_{_{O_2}}) \phi_{j,1} =- \beta_2 (\lambda_j p_{_{O_2}}) \phi_{j,2}, \label{boundary1} \\
\nabla \phi_{j,2} \cdot \vec{n} + \alpha_2 (\lambda_j p_{_{O_2}}) \phi_{j,2} =- \beta_1 (\lambda_j p_{_{O_2}}) \phi_{j,1}, \label{boundary2}
\end{align}
where $\vec{n}$ is the outward unit normal vector, $\alpha_n = \frac{5}{96} \left( 34 + (-1)^{n-1}11 \sqrt{\frac{6}{5}} \right)$ and $\beta_n = \frac{5}{96} \left( 2 + (-1)^n \sqrt{\frac{6}{5}} \right)$ ($n=1,2$). 

Eqs. \eqref{sp3sol} to \eqref{boundary2} define an approximate method to calculate \eqref{rteSph}, and we denote this method as SP$_3$ Larsen BC method.

\subsection{Fast multipole method for the evaluation of integral}\label{sec:fmm}
Instead of utilizing differential equations for approximation, we evaluate the numerical quadrature of \eqref{integral} efficiently in \cite{Lin2020psst} based on the fast multipole method \cite{ying2004}. Compared with the previous PDE-based approximations, this new evaluation method is more accurate and robust \cite{Lin2020psst}, and is free of fitting additional coefficients. Generally, the quadrature of \eqref{integral} can be written as
\begin{equation}
S_{\rm ph}(\vec{x}_i) = \sum_{j=1}^{N_{\rm pt}} G(\vec{x}_i ,\vec{y}_j) I_h(\vec{y}_j), \qquad i=1,\cdots,N_{\rm pt},
\label{summation}
\end{equation}
where $G(\cdot,\cdot)$ is the discrete kernel function, and $I_h(\vec{y}_j)$ is the product of \eqref{Ifun} and the mesh size at $\vec{y}_j$. Interested readers may refer \cite{Lin2020psst} for more details.

The fast multipole method  evaluates \eqref{summation} based on a hierarchical tree (octree in 3D)\cite{ying2004}. This hierarchical tree is constructed such that each leaf cube contains no more than a prescribed number of points. In the uniform case where a uniform grid is adopted for discrete points, for each target point $\vec{x}_i$ in a leaf cube $B$, the FMM partitions the summation \eqref{summation} into near interactions $\mathcal{N}(B)$ and far interactions $\mathcal{F}(B)$ as
\begin{equation}
S_{\rm ph}(\vec{x}_i) = \sum_{\vec{y}_j \in \mathcal{N}(B)} G(\vec{x}_i ,\vec{y}_j) I_h(\vec{y}_j) + \sum_{\vec{y}_j \in \mathcal{F}(B)} G(\vec{x}_i ,\vec{y}_j) I_h(\vec{y}_j),
\label{nearfar}
\end{equation}
where the near interactions are calculated exactly, and the far interactions are approximated using the hierarchy tree because of the low-rankness of kernel $G(\cdot, \cdot)$.

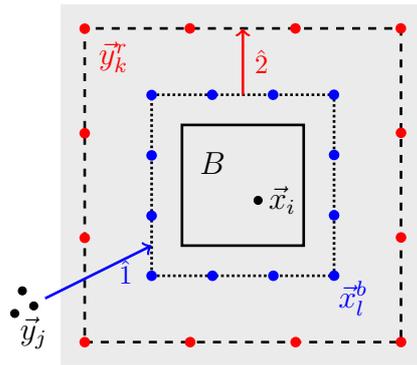
\begin{figure}[htb!]
\centering
\begin{tikzpicture}
\fill[black!8!white] (4.6,-2.4) rectangle (9.4,2.4);
\draw[line width = 1pt] (6.2,-0.8) rectangle (7.8,0.8);
\node[draw=black!8!white, circle, inner sep =0pt] at (6.6,0.3) {$B$};
\filldraw[black] (7.2,-0.2) circle (1.5pt) node[anchor=west]{$\vec{x}_i$};
\draw[densely dotted, line width=1pt] (5.8,-1.2) rectangle (8.2,1.2);
\draw[dashed, line width = 1pt] (4.92,-2.08) rectangle (9.08,2.08);
\draw [blue, line width=4pt,line cap=round,dash pattern=on 0pt off 0.8cm] (5.8,-1.2) rectangle (8.2,1.2);
\draw [red, line width=4pt,line cap=round,dash pattern=on 0pt off 1.38667cm] (4.92,-2.08) rectangle (9.08,2.08);
\filldraw[black] (4.25,-1.6) circle (1.5pt) node[anchor=north]{$\vec{y}_j$};
\filldraw[black] (4.1, -1.4) circle (1.5pt);
\filldraw[black] (4.0, -1.7) circle (1.5pt);
\node[circle, inner sep =0pt, text=red] at (5.3,1.7) {$\vec{y}^{r}_k$};
\node[circle, inner sep =0pt, text=blue] at (8.45,-1.5) {$\vec{x}^{b}_l$};
\draw[->, line width = 1pt, blue] (4.4,-1.5) -- (5.8,-0.8) node[midway, right] {{\footnotesize $\ \hat{1}$}};
\draw[->, line width = 1pt, red] (7,1.2) -- (7,2.08) node[midway, right] {{\footnotesize $\hat{2}$}};
\end{tikzpicture}
\caption{\label{fmulloc}Cross section of source and checking points of box $B$. Red dots connected by dashed lines denote artificial source points, and blue dots connected by dotted lines denote checking points. Shadow part is the near range of $B$. Arrow $\hat{1}$ is the summation in the right-hand side of \eqref{s2l}, and arrow $\hat{2}$ is the calculation of $I_a(\vec{y}_k^{r})$.}
\end{figure}

The far interactions $\mathcal{F}(B)$ in \eqref{nearfar} is approximated by a small number of artificial source points enclosing $B$. This idea can be explained in figure \ref{fmulloc}, where the interaction from the source points $\vec{y}_j$ is approximated through the red artificial source points enclosing $B$. The blue points serve as checking points which are locations to check the artificial interaction from the red artificial source points is identical to the interaction from the original source points $\vec{y}_j$. If we denote the red source points as $\vec{y}_{k}^{r}$ and the corresponding artificial intensity at these points as $I_{a}(\vec{y}_k^{r})$, then $I_{a}(\vec{y}_k^{r})$ can be obtained by solving the following linear equations
\begin{equation}
\sum_{\vec{y}_k^{r}} G(\vec{x}_l^{b},\vec{y}_k^{r}) I_a(\vec{y}_k^{r})= \sum_{\vec{y}_j \in \mathcal{F}(B)} G(\vec{x}_l^{b},\vec{y}_j) I_h(\vec{y}_j),\ \forall \vec{x}_l^{b},
\label{s2l}
\end{equation}
where $\vec{x}_l^{b}$ are the blue checking points. Then, the far interaction in \eqref{nearfar} could be approximated by a small number of points $\vec{y}_k^{r}$ and their artificial intensity $I_{a}(\vec{y}_k^{r})$.

The hierarchy tree is used to calculate the right-hand side of \eqref{s2l} approximately. The idea is similar to \eqref{s2l}, which is constructing artificial source points and corresponding artificial intensities for other boxes in the hierarchy tree. Before obtaining $I_{a}(\vec{y}_k^{B,d})$, we first calculate the artificial intensities of the parent box of $B$, and then those original source points located at the far interactions of the parent box could be approximated by the artificial intensities of the parent box of $B$. This procedure could be done recursively following the hierarchy tree. Detailed illustration and the outline of whole algorithm could be found in \cite{ying2004, Lin2020psst}.

\section{Adaptive strategies in the fast multipole method}\label{fmm}
As stated in Section \ref{sec:fmm}, the fast multipole method does not need fitting additional coefficients, and performs more accurate and robust than the two PDE-based methods. However, under some typical settings (e.g., figure 9 in \cite{Lin2020psst}), its computational cost is a little larger than the PDE-based methods with efficient boundaries, though their costs are at the same order. 
In some circumstances, we might not require very high accuracy but wish to gain more efficiency. This motivates us to further speed up the calculation of the fast multipole method with acceptable sacrifice on the accuracy.

We introduce three adaptive approaches to calculate \eqref{summation} more efficiently. The idea of these approaches is to pick up those ``important'' points $\vec{x}_i$ and $\vec{y}_j$ in the summation \eqref{summation} instead of summing up all $N_{pt}$ points. In other words, we take out those $\vec{y}_j$ which are less ``important'' to the summation and those $\vec{x}_i$ which are less ``important'' to observe. Similar idea was used in the calculation of \eqref{integral} using the integral form to ease the expensive computation. In \cite{kulikovsky2000}, the author made a truncation based on $|\vec{x}_i-\vec{y}_j|$, which picked up some nearest points in the summation. A coarser grid was introduced in \cite{Hallac2003, Georghiou2005} to compute the integral and then the interpolation was taken to get the photoionization rate on the original grid, which could be regarded as picking $\vec{y}_j$ on a coarser grid. The stochastic version of \eqref{integral} in \cite{bagheri2019effect, chanrion2008} can also be viewed as picking up $\vec{y}_j$ by a prescribed distribution.

Our first approach is to pick $\vec{y}_j$ based on the intensity. The function $I(\vec{y})$ in \eqref{Ifun} has a fast decay when $\vec{y}$ is away from the main channel of a streamer. In such case, the magnitude of $I_h(\vec{y}_j)$ in \eqref{summation} is relatively small and therefore the summation from these $I_h(\vec{y}_j)$ is neglectable to the photoionization rate $S_{\rm ph}(\vec{x}_i)$. As a result, we could pick those $\vec{y}_j$ with high intensities and rewrite the summation \eqref{summation} as
\begin{equation}
S_{\rm ph}(\vec{x}_i) \approx \sum_{I_h(\vec{y}_j) \geq I_0} G(\vec{x}_i ,\vec{y}_j) I_h(\vec{y}_j), \qquad i=1,\cdots,N_{\rm pt},
\label{summationI}
\end{equation}
where $I_0$ is a threshold of the intensity. This approach could also be regarded as neglecting those source radiation with less intensities.  

The second approach is a truncation of the simulation domain for $S_{\rm ph}$, which is the selection of $\vec{x}_i$ in \eqref{summationI}. This approach makes use of the fact that it is not necessary to compute all $\vec{x}_i$ under some circumstances. In the fluid model of streamer discharges, photoionization is partially used to provide seed electrons in front of the head of a streamer. The effect of photoionization is not essential for the region where the collision ionization dominates the ionization and the photoionization has a much weaker effect. Therefore, only a portion of $\vec{x}_i$ is essential, and we can just pick these $\vec{x}_i$. However, the selection of $\vec{x}_i$ is more problem-oriented, and it is hard to give a general criterion; see sections \ref{comparison} and \ref{comparison1} for some criteria in different numerical examples. 

\begin{figure}[htb!]
\centering
\begin{tikzpicture}[scale=0.8]
\fill[red!20!white] (0,0) rectangle (6,6);
\fill[blue!40!white] (5,2) rectangle (6,6);
\fill[blue!40!white] (4,4.5) rectangle (5,6);
\fill[blue!40!white] (4.5,4) rectangle (5,4.5);
\fill[white] (2,2) rectangle (4,4);
\draw [step=2] (0,0) grid (6,6);
\draw [step=1] (4,2) grid (6,6);
\draw [step=0.5] (4,4) grid (5,5);
\node[draw=white, circle, inner sep =0pt] at (3,3) {$B$};
\end{tikzpicture}
\caption{\label{fig:w}Cross section of boxes in $\mathcal{N}(B)$. The red-shadow boxes are directly adjacent to $B$, and the blue-shadow boxes are not adjacent to $B$ but in $\mathcal{N}(B)$.}
\end{figure}
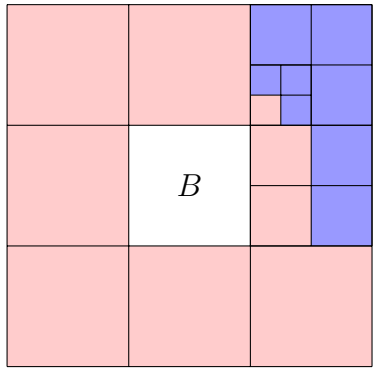

The third approach is applied to the algorithm of the fast multipole method to skip some calculation of $\vec{y}_j \in \mathcal{N}(B)$ in \eqref{nearfar}. After accumulating $\vec{y}_j \in \mathcal{F}(B)$, $\vec{y}_j \in B$ and those $\vec{y}_j$ located in the boxes directly adjacent to $B$, we get an estimate of the magnitude of $S_{\rm ph}(\vec{x}_i)$ in \eqref{nearfar}. If the ratio of the remaining $I_h(\vec{y}_j)$ to the estimated $S_{\rm ph}(\vec{x}_i)$ is less than a prescribed number $r_s$, we skip these $I_h(\vec{y}_j)$ in the summation. Figure \ref{fig:w} gives an example of boxes in $\mathcal{N}(B)$ but not directly adjacent to $B$. As the figure shows, this approach sacrifices the accuracy of those $S_{\rm ph}(\vec{x}_i)$ in the box surrounded by multi-level boxes, which would happen near the boundary of selected points of the first two approaches. In the case of a streamer discharge, this boundary is either dominated by the impact ionization or far away from the channel of streamer where the accuracy of photoionization rate is less important. In this paper, this approach will only be applied in the simulation of streamer discharges in section \ref{comparison1} with $r_s = 10^{-4}$.

\section{Results and comparison for computing photoionization rate}\label{comparison}
In this section, we apply different methods in sections \ref{model} and \ref{fmm} to calculate the photoionization rate $S_{\rm ph}$ in \eqref{integral}. These methods are compared in terms of accuracy and efficiency, and the notations of them are summarized in table \ref{diffnames}.

\Table{\label{diffnames}Notations of several methods introduced in this paper.}
\br
\ns
Notation of method & Brief description \\
\mr
Classical Int & Direct calculation on (\ref{integral}), with (\ref{Ifun}) and (\ref{gfun}) \\
FMM classical Int & Fast multipole method based on (\ref{integral}), with (\ref{Ifun}) and (\ref{gfun}) \\
FMM Int Src & FMM classical Int method with adaptive source points based on \eqref{summationI} \\
FMM Int SrcTrg & FMM Int Src method with only a portion of $\vec{x}_i$ in \eqref{summationI} \\
Helmholtz zero BC & Three term summation on \eqref{sphHel}, by solving \eqref{Hel} with zero BC\\
SP$_3$ Larsen BC & Three-group summation on \eqref{rteSph}, by solving \eqref{phi1} and \eqref{phi2} with BCs \eqref{boundary1} and \eqref{boundary2} \\
\br
\end{tabular}
\end{indented}
\end{table}

All the numerical simulations in this paper were executed via the MPI parallelism on the Tianhe2-JK cluster located at Beijing Computational Science Research Center. We adopted 4 nodes (20 cores in each node) for the computations in this section. 

The computational domain was taken as $V=V'=[0, x_d] \times [0, y_d] \times [0,z_d]$\, cm$^3$, and its center was denoted as $\vec{x}_0=(x_0,y_0,z_0)^T= (x_d/2, y_d/2, z_d/2)^T$\,cm. $V$ was partitioned by a uniform mesh with $n_x \times n_y \times n_z$ cells. 

Following \cite{Lin2020psst}, the global relative error $\mathcal{E}_V$ and the local relative error $\mathcal{E}_{\delta}(\vec{x}_0)$ are used to quantify the accuracy of different numerical methods:
\begin{equation}
\begin{split}
&\mathcal{E}_V := \frac{ \| S_{\rm ph}^{\rm num}(\vec{x})-S_{\rm ph}^{\rm ref}(\vec{x}) \|_2}{\|S_{\rm ph}^{\rm ref}(\vec{x})\|_2}\times 100\%, \\
&\mathcal{E}_{\delta}(\vec{x}_0):= \frac{1}{N_{\text{tot}}} \sum_{|\vec{x} - \vec{x}_0 | \leq \delta} \frac{| S_{\rm ph}^{\rm num}(\vec{x})-S_{\rm ph}^{\rm ref}(\vec{x})|}{S_{\rm ph}^{\rm ref}(\vec{x})}\times 100\%,\\
\end{split}
\label{relativefun}
\end{equation}
where $\|\cdot\|_2$ is the discrete $L^2$-norm, $\delta>0$ is a constant indicating the radius of a ball centered at $\vec{x}_0$, and $N_{\text{tot}}$ is the number of grid points within this ball. $S_{\rm ph}^{\rm ref}(\vec{x})$ is obtained by the (discrete) Classical Int method which serves as the reference, and $S_{\rm ph}^{\rm num}(\vec{x})$ is calculated by a given numerical method.

To solve the elliptic equations (in Helmholtz zero BC method and SP$_3$ Larsen BC method) efficiently, we adopted the multigrid-preconditioned FGMRES solver in the numerical experiments; Interested readers could refer to \cite{lin2018} for a detailed illustration of its performance.

\subsection{Gaussian emission source with different sizes of the domain}\label{DiffR}

This example follows \cite{Lin2020psst, bourdon2007} to compute the photoionization generated from a single Gaussian emission source as
\begin{equation}
S_{i}(\vec{x}) = 1.53 \times 10^{25} \exp\left(- \left( (x-x_0)^2 + (y-y_0)^2 + (z-z_0)^2\right) / \sigma^2 \right) \text{cm}^{-3}\,\text{s}^{-1},
\label{onegaussiansrc}
\end{equation}
where $\sigma>0$ is a constant to be given later.
The other parameters in \eqref{integral}-\eqref{gfun} are chosen as \cite{bourdon2007,segur2006}: $p_q =30$\,Torr, $p=760$\,Torr, $\xi=0.1$, $\omega/ \alpha = 0.6$, $p_{_{O_2}}=150$\,Torr; $\delta=5\sigma$.

$n_x=n_y=320$ and $n_z=160$ were taken in this example. Under this grid size with mentioned parameters, the reference solution of the Classical Int method was computed in \cite{Lin2020psst}, and we reuse the data of the reference solution to avoid redundant computation.

In this example, we take the adaptive strategy for FMM Int Src method in \eqref{summationI} as $I_0  = 0.1 \% I_{\max}$, where $I_{\max}$ is the maximum value of $I_h(\vec{y}_k)$ for all $\vec{y}_k$ in domain $V$. This strategy drops the source points whose magnitude of radiation is less than $0.1 \% I_{\max}$. 

\begin{figure}[htb!]
\centering
\includegraphics[width=0.45\textwidth]{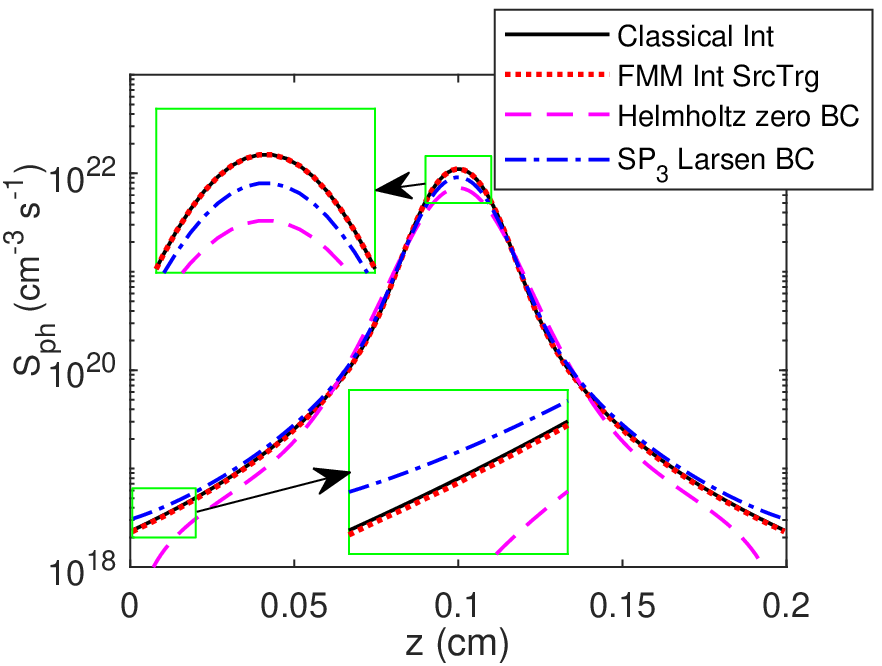}
\includegraphics[width=0.45\textwidth]{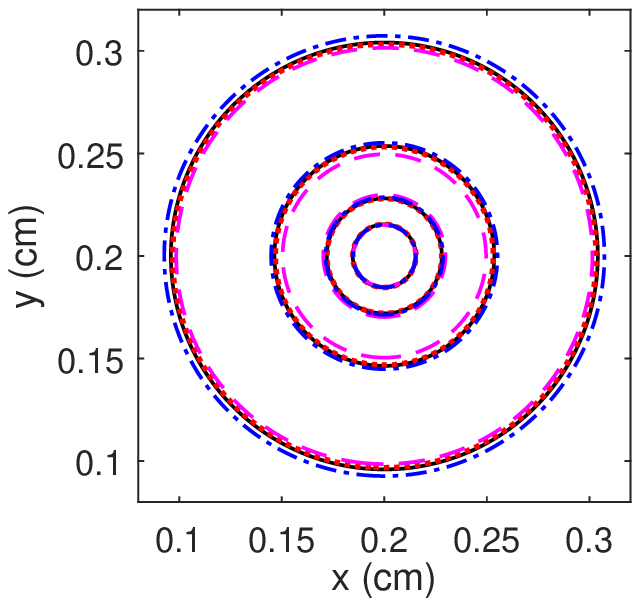}
\caption{\label{fonegauss001}Photoionization rate $S_{\rm ph}$ calculated from one Gaussian source \eqref{onegaussiansrc}. $x_d=y_d=0.4$\,cm, $z_d=0.2$\,cm, $\sigma=0.01$\,cm. The left sub-figure is $S_{\rm ph}$ along line $x=y=0.2$\,cm; the right sub-figure is the contours of $S_{\rm ph} = 2 \times 10^{18}$, $2 \times 10^{19}$, $2 \times 10^{20}$, $2 \times 10^{21}$\,cm$^{-3}$\,s$^{-1}$ on the plane $z=0.1$\,cm. The color and format of lines in the right sub-figure are identical to those in the left sub-figure.}
\end{figure}

\begin{figure}[htb!]
\centering
\includegraphics[width=0.45\textwidth]{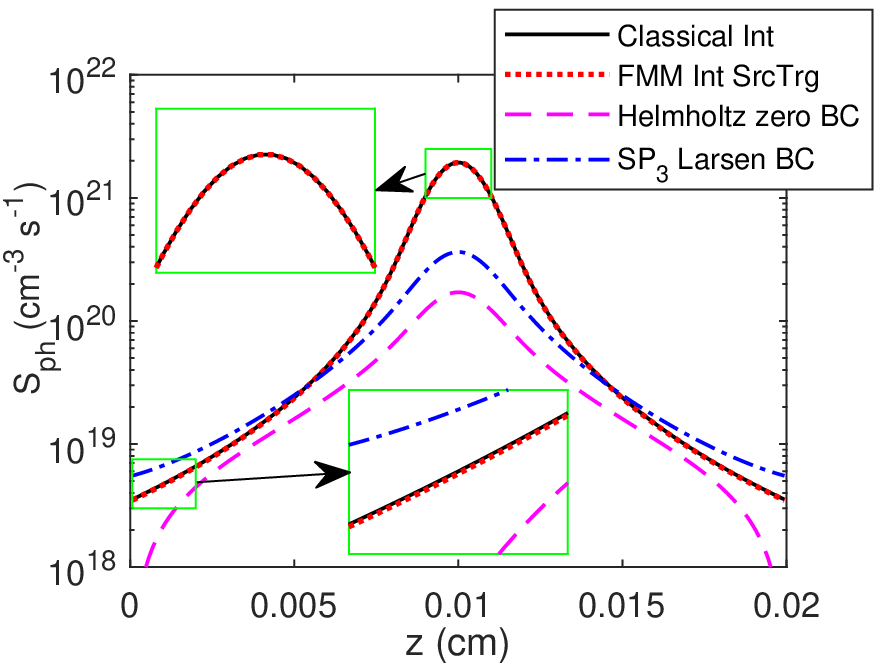}
\includegraphics[width=0.45\textwidth]{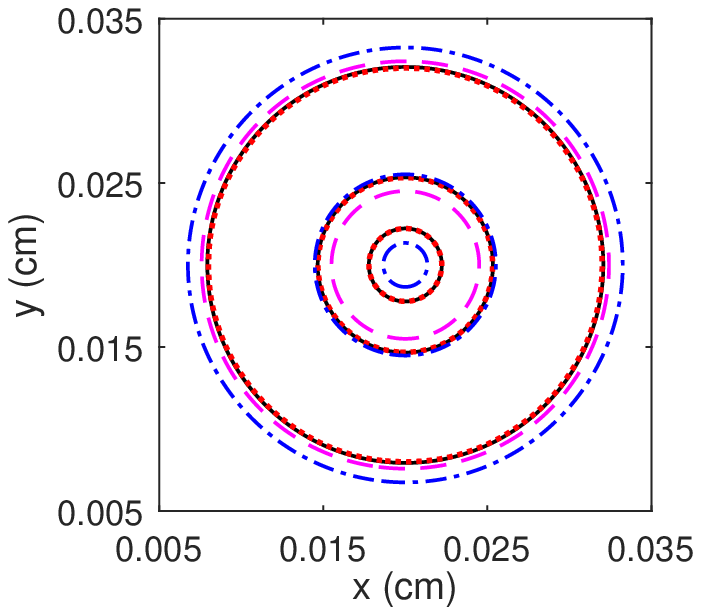}
\caption{\label{fonegauss0001}Photoionization rate $S_{\rm ph}$ calculated from one Gaussian source in (\ref{onegaussiansrc}). $x_d=y_d=0.04$\,cm, $z_d=0.02$\,cm, $\sigma=0.001$\,cm. The left sub-figure is $S_{\rm ph}$ along line $x=y=0.2$\,cm; the right sub-figure is the contours of $S_{\rm ph} = 2 \times 10^{18}$, $2 \times 10^{19}$, $2 \times 10^{20}$ on the plane $z=0.1$\,cm. The color and format of lines in the right sub-figure are identical to those in the left sub-figure.}
\end{figure}

As stated in the second approach in section \ref{fmm}, the adaptive strategy for selecting $\vec{x}_i$ in the FMM Int SrcTrg method is demand-oriented. In this section, the main purpose is to compare the accuracy and efficiency of different methods when calculating the photoionization rate. Besides comparing the error in the whole domain, we also illustrate the results in figures \ref{fonegauss001} and \ref{fonegauss0001} which are contours in a cross section and a line. Therefore, in this case, we regard plotting these two figures as the oriented demand in the FMM Int SrcTrg method, and those $\vec{x}_i$ needed for plotting the figures are chosen as the target points. For simplicity, we call this strategy as demand-oriented strategy hereafter in section \ref{comparison}. The demand-oriented strategy in this example chooses about 0.2 million points $\vec{x}_i$ around a cross section and a line.


Similar to \cite{bourdon2007}, we consider two different sizes of the domain $V$:
\begin{enumerate}
\item $x_d=y_d=0.4$\,cm, $z_d=0.2$\,cm, $\sigma=0.01$\,cm;
\item $x_d=y_d=0.04$\,cm, $z_d=0.02$\,cm, $\sigma=0.001$\,cm.
\end{enumerate}
The numerical results are plotted in figures \ref{fonegauss001} and \ref{fonegauss0001}, and the relative errors are computed in tables \ref{tonegaussiansrc001} and \ref{tonegaussiansrc0001}.

\Table{\label{tonegaussiansrc001}Time usage and relative error of methods indicated in Table \ref{diffnames}, for the case of single Gaussian source $x_d=y_d=0.4$\,cm, $z_d=0.2$\,cm, $\sigma=0.01$\,cm. $\vec{x}_0 = (0.2, 0.2, 0.1)^T$\,cm and $\delta = 5 \sigma$.}
\br
Method & Time usage (s) & $\mathcal{E}_V$ & $\mathcal{E}_{\delta}(\vec{x}_0)$\\
\br
Classical Int & 184248$^{\rm a}$ & --- & --- \\
FMM classical Int & 28.5169 & 0.21\% & 1.30\% \\
FMM Int Src (0.1\%) & 8.32157 & 0.32\% & 1.94\% \\
FMM Int SrcTrg & 1.04643 & N.A.$^{\rm b}$ & N.A.$^{\rm b}$ \\
Helmholtz zero BC & 3.91421 & 25.33\% & 16.37\% \\
SP$_3$ Larsen BC & 11.9803 & 12.05\% & 8.49\% \\
\br
\end{tabular}
\item[] $^{\rm a}$Datum is taken from \cite{Lin2020psst} under same computational setting. 
\item[] $^{\rm b}$Not fully defined. 
\end{indented}
\end{table}

Figures \ref{fonegauss001} and \ref{fonegauss0001} clearly illustrate that the FMM Int SrcTrg method gives more accurate results compared to the Helmholtz zero BC method and the SP$_3$ Larsen method. The difference between the lines of reference Classical Int method and lines of the FMM SrcTrg method is negligible, while the deviation of other two methods can be clearly observed, especially in the central region and near the boundary. This observation coincides with the conclusion in the same example in \cite{Lin2020psst}, where the FMM method is superior in terms of accuracy. The FMM Int SrcTrg method, with given adaptive strategies, maintains the high accuracy of the FMM method and still gives good approximation to the reference solution.


The accuracy of the FMM Int Src method are shown quantitatively in tables \ref{tonegaussiansrc001} and \ref{tonegaussiansrc0001}. Similar to the results of the FMM classical Int method, the relative error of FMM Int Src method is at least one order magnitude less than the Helmholtz method and SP$_3$ method. As an adaptive version of FMM classical Int method, the FMM Int Src method preserves the accuracy properly.  

The efficiency of the FMM Int Src method and the FMM Int SrcTrg method is more interesting, which could be seen in tables \ref{tonegaussiansrc001} and \ref{tonegaussiansrc0001}. With the inclusion of first adaptive strategy ($I_0=0.1\%I_{\max}$), the time usage of the FMM Int Src method becomes 1/3 of the FMM classical Int method, and it is even less than that of the SP$_3$ method. Furthermore, the demand-oriented FMM Int SrcTrg method accelerates the FMM Int Src method significantly, and it becomes the fastest method which takes only 1/4 of the time cost of the Helmholtz method. Although the adaptive FMM method is slightly less accurate than the FMM classical Int method, the reduction of computational cost make it comparable to the Helmholtz method and SP$_3$ method in terms of efficiency. Together with its performance in accuracy, the adaptive FMM method shows its outstanding competitiveness for computing the photoionization rates.

\Table{\label{tonegaussiansrc0001}Time usage and relative error of methods indicated in Table \ref{diffnames}, for the case of one Gaussian source $x_d=y_d=0.04$\,cm, $z_d=0.02$\,cm, $\sigma=0.001$\,cm. $\vec{x}_0 = (0.02, 0.02, 0.01)^T$\,cm and $\delta = 5 \sigma$.} 
\br
Method & Time usage (s) & $\mathcal{E}_V$ & $\mathcal{E}_{\delta}(\vec{x}_0)$\\
\br
Classical Int & 183761$^{\rm a}$ & --- & --- \\
FMM classical Int & 28.5142 & 0.22\% & 0.53\% \\
FMM Int Src (0.1\%) & 8.44722 & 0.39\% & 0.94\% \\
FMM Int SrcTrg & 1.10534 & N.A.$^{\rm b}$ & N.A.$^{\rm b}$ \\
Helmholtz zero BC & 4.26866 & 83.26\% & 48.03\% \\
SP$_3$ Larsen BC & 16.8232 & 68.92\% & 16.67\% \\
\br
\end{tabular}
\item[] $^{\rm a}$Datum is taken from \cite{Lin2020psst} under same computational setting.
\item[] $^{\rm b}$Not fully defined. 
\end{indented}
\end{table}

\subsection{Gaussian emission source with different pressures}\label{DiffP}
The robustness of different methods (including the PDE-based methods and the FMM classical Int method) with respect to partial pressure $p_{_{O_2}}$ was discussed in \cite{Lin2020psst}. The results showed that the FMM classical method always exhibits accurate simulation for different pressures using similar computational cost, while the PDE-based methods becomes less accurate and slightly more time-consuming as the pressure decreases. In this subsection, we will adopt the same example in \cite{Lin2020psst} to test the robustness of the fast multipole method when adaptive strategies are included.

As a reference, the setting of the example in \cite{Lin2020psst} is listed here. The photoionization rate $S_{\rm ph}(\vec{x})$ in \eqref{integral} is computed and compared when $I(\cdot)$ in \eqref{Ifun} is taken as
\begin{equation}
I(\vec{x}) = 4 \pi \xi c \exp \left[ - \frac{ (x-x_0)^2 + (y-y_0)^2+(z-z_0)^2}{\sigma^2} \right] \text{cm}^{-3}\,\text{s}^{-1},
\label{po2init}
\end{equation}
where constant $c$ is the speed of light, $\xi=0.1$, $\sigma=0.01$\,cm and $V=[0,0.25] \times [0,0.25] \times [0,1.4]$\,cm$^{3}$. Furthermore, the ratio of $p_{_{O_2}}$ and the air pressure is fixed as $p_{_{O_2}}/p=0.2$. The domain $V$ is partitioned by $256 \times 256 \times 320$ cells. We take $\delta=5\sigma$ in $\mathcal{E}_{\delta}(\vec{x}_0)$.

\begin{figure}[htb!]
\centering
\includegraphics[width=0.45\textwidth]{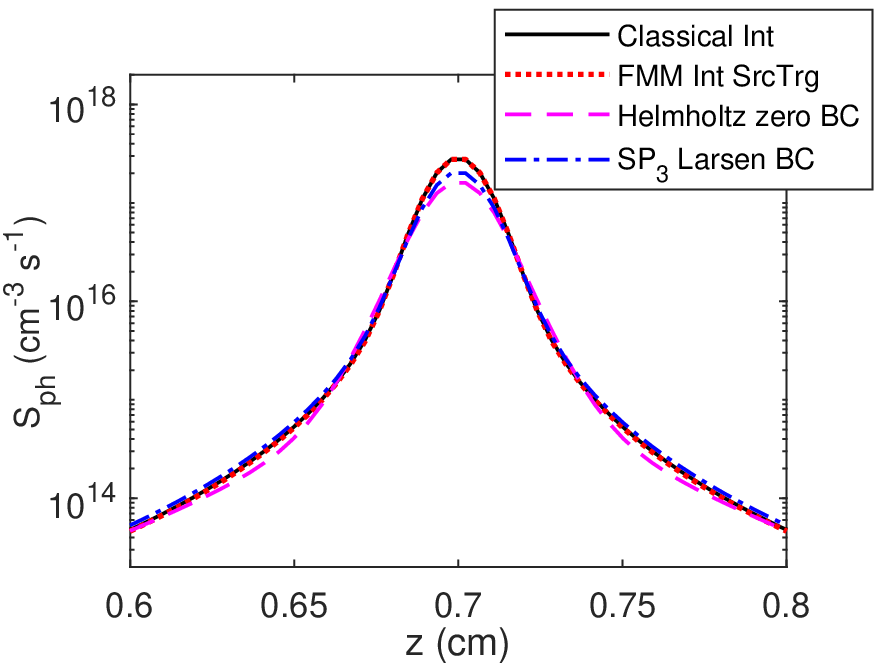}
\includegraphics[width=0.45\textwidth]{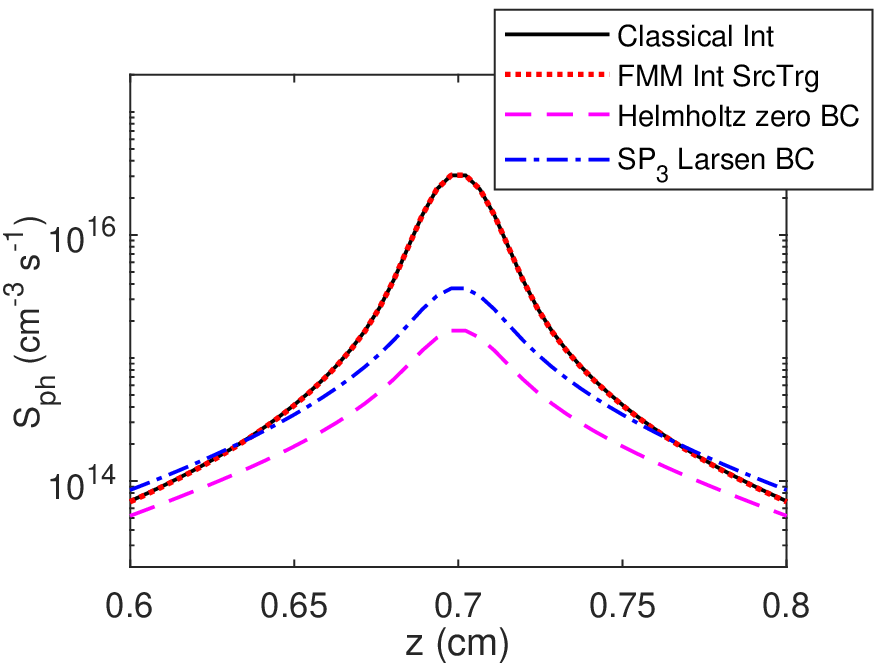}
\caption{\label{fdiffpi}Photoionization rate $S_{\rm ph}$ along line $x=y=0.125$\,cm, calculated from one Gaussian in (\ref{po2init}) with: $p_{_{O_2}}=160$\,Torr (left sub-figure); $p_{_{O_2}}=10$\,Torr (right sub-figure).}
\end{figure}

Two different pressures are considered, namely, $p_{_{O_2}}=160$\,Torr and $p_{_{O_2}}=10$\,Torr. Figure \ref{fdiffpi} shows the photoionization rate along the central vertical line of the domain, and the computational cost and the numerical error are summarized in tables \ref{tdiffpi} and \ref{tdiffpiii}.

The adaptive strategy is taken similar to the one in section \ref{DiffR}. For the FMM Int Src method, the adaptivity is to choose those $\vec{y}_j$ such that $I(\vec{y}_j) \geq 0.1 \% I_{\max}$. For the FMM Int SrcTrg method, since figure \ref{fdiffpi} only plots results on the central vertical line, only those $\vec{x}_i$ adjacent to this line are selected. 

\Table{\label{tdiffpi}Time usage and relative error of methods in Table \ref{diffnames}, for the case of one Gaussian in (\ref{po2init}) with $p_{_{O_2}}=160$\,Torr. $\vec{x}_0$ is the center of $V$ and $\delta = 0.05$\,cm.} 
\br
Method & Time usage (s) &$\mathcal{E}_V$ &$\mathcal{E}_{\delta}(\vec{x}_0)$ \\
\br
Classical Int & 292196$^{\rm a}$ & --- & --- \\
FMM classical Int & 26.9083 & 0.15\% & 1.24\% \\
FMM Int Src ($0.1\%$) & 8.38895 & 0.25\% & 1.89\% \\
FMM Int SrcTrg & 0.456448 & N.A.$^{\rm b}$ & N.A.$^{\rm b}$ \\
Helmholtz zero BC & 5.92624 & 31.93\% & 16.49\% \\
SP$_3$ Larsen BC & 17.8230 & 21.31\% & 8.74\% \\
\br
\end{tabular}
\item[] $^{\rm a}$Datum is taken from \cite{Lin2020psst} under same computational setting.
\item[] $^{\rm b}$Not fully defined.
\end{indented}
\end{table}

\Table{\label{tdiffpiii}Time usage and relative error of methods in Table \ref{diffnames}, for the case of one Gaussian in (\ref{po2init}) with $p_{_{O_2}}=10$\,Torr. $\vec{x}_0$ is the center of $V$ and $\delta = 0.05$\,cm.} 
\br
Method & Time usage (s) &$\mathcal{E}_V$ &$\mathcal{E}_{\delta}(\vec{x}_0)$\\
\br
Classical Int & 292426$^{\rm a}$ & --- & --- \\
FMM classical Int & 26.9367 & 0.19\% & 0.44\% \\
FMM Int Src ($0.1\%$) & 8.43274 & 0.35\% & 0.82\% \\
FMM Int SrcTrg & 0.486245 & N.A.$^{\rm b}$ & N.A.$^{\rm b}$ \\
Helmholtz zero BC & 6.63734 & 88.73\% & 65.87\% \\
SP$_3$ Larsen BC & 23.8077 & 77.74\% & 35.11\% \\
\br
\end{tabular}
\item[] $^{\rm a}$Datum is taken from \cite{Lin2020psst} under same computational setting.
\item[] $^{\rm b}$Not fully defined.
\end{indented}
\end{table}

Figure \ref{fdiffpi} shows the accuracy of the adaptive FMM is stable for the two pressures. The curves of the FMM Int SrcTrg method match the curves of Classical Int method very closely in both high and low pressure cases. On the contrary, the PDE-based methods give deviated results when the pressure becomes lower. This phenomenon can also be observed in tables \ref{tdiffpi} and \ref{tdiffpiii} quantitatively. The FMM Int Src method retains high accuracy at the order of $1\%$ in both pressures, which is comparable to the accuracy of the FMM classical Int method. More importantly, the efficiency of FMM is greatly enhanced with the inclusion of adaptive strategies. In this specific example, only few points along the single line are required, therefore, it leads to the tiny time usage of the FMM Int SrcTrg method, which is less than $1/10$ of the second efficient method (i.e., Helmholtz zero BC method) in both two pressures.

\begin{figure}[htb!]
\centering
\includegraphics[width=0.48\textwidth]{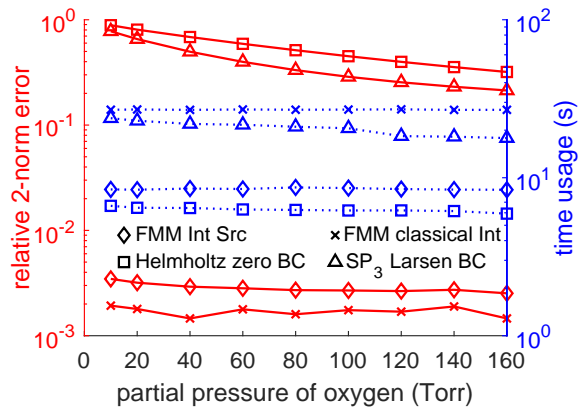}
\caption{\label{pressureCom}Error $\mathcal{E}_V$ (red color, solid line) and time usages (blue color, dotted line) of FMM Int Src ($0.1\%$), FMM classical Int, Helmholtz zero BC and SP$_3$ Larsen BC methods with different partial pressures of oxygen.}
\end{figure}

To observe more details about the error and computational time when the pressure varies, we follow \cite{Lin2020psst} to maintain the same setting except changing $p_{_{O_2}}$ to be some other values between $10$\,Torr and $160$\,Torr. The FMM Int Src method is compared with the FMM classical Int, Helmholtz zero BC and SP$_3$ Larsen BC methods together in figure \ref{pressureCom}. 

It could be seen that the FMM classical Int method provides the most accurate results with the price of larger computational cost for all pressures. With the adaptive strategy, its error becomes larger, but the error of the FMM Int Src method is at the same order of the FMM classical Int method. Although the error of the FMM Int Src method increases slightly when the pressure decreases, it is still roughly two order magnitudes less than other two PDE-based methods for all pressures. On the other hand, the time usage of the FMM Int Src method is independent of pressure, and is close to that of the Helmholtz zero BC method. Therefore, the FMM Int Src method has comparable accuracy to the most accurate method and comparable computational cost to the fastest method, and this comparability has little dependency on pressure.


\subsection{Gaussian emission source with different adaptive strategies}\label{chapteradaptive}
Previous examples take the same adaptive strategy for choosing source points in the FMM Int Src method and the FMM Int SrcTrg method, which chooses those $\vec{y}_j$ such that $I(\vec{y}_j) \geq 0.1 \% I_{\max}$. It could be expected that there is a speed-or-accuracy trade-off in the adaptive strategy, namely, more points yields better accuracy but requires more computational costs. In this subsection, we will focus on adaptive strategies in the form
\begin{equation}\label{eq:sourcepct}
I(\vec{y}_j) \geq I_{pct} I_{\max}   
\end{equation}
and discuss the effect of different percentage $I_{pct}$.

For convenience, we reuse the setting of examples in sections \ref{DiffR} and \ref{DiffP}. All the parameters are set identical to the previous examples, except changing the adaptive strategy of choosing source points from $I(\vec{y}_j) \geq 0.1 \% I_{\max}$ to \eqref{eq:sourcepct}.

We reuse the two examples (two different sizes of the domain where $z_d = 0.2$, $0.02$\,cm) in Section \ref{DiffR}. The results are summarized in tables \ref{tadaptiveR1} and \ref{tadaptiveR2}, where the speed-or-accuracy trade-off can be clearly observed in both domains regardless of their sizes. On the one hand, if we only require the relative error to be around $20\%$ ($I_{pct} = 10\%$), the time usage of the FMM Int Src method becomes only $1/8$ of that of the FMM classical Int method in tables \ref{tonegaussiansrc001} and \ref{tonegaussiansrc0001}, and even less than that of the Helmholtz zero BC method. If we wish to maintain the accuracy of FMM method and keep the relative error at the order of $1\%$, we should take $I_{pct} = 0.1\%$ to ensure that sufficient source points are selected. 
\Table{\label{tadaptiveR1}Time usage and relative error of FMM Int Src method with different $I_{pct}$ in the adaptive strategy \eqref{eq:sourcepct}, for the case of single Gaussian source in \eqref{onegaussiansrc} with $x_d=y_d=0.4$\,cm, $z_d=0.2$\,cm, $\sigma=0.01$\,cm. $\vec{x}_0 = (0.2, 0.2, 0.1)^T$\,cm and $\delta = 5 \sigma$.}
\br
Method & Time usage (s) & $\mathcal{E}_V$ & $\mathcal{E}_{\delta}(\vec{x}_0)$\\
\br
FMM Int Src (0.01\%) & 12.1093 & 0.21\% & 1.35\% \\
FMM Int Src (0.1\%) & 8.32157 & 0.32\% & 1.94\% \\
FMM Int Src (1\%) & 5.58410 & 1.58\% & 5.70\% \\
FMM Int Src (10\%) & 3.35450 & 13.30\% & 26.12\% \\
\br
\end{tabular}
\end{indented}
\end{table}

\Table{\label{tadaptiveR2}Time usage and relative error of FMM Int Src method with different $I_{pct}$ in the adaptive strategy \eqref{eq:sourcepct}, for the case of one Gaussian source in \eqref{onegaussiansrc} with $x_d=y_d=0.04$\,cm, $z_d=0.02$\,cm, $\sigma=0.001$\,cm. $\vec{x}_0 = (0.02, 0.02, 0.01)^T$\,cm and $\delta = 5 \sigma$.} 
\br
Method & Time usage (s) & $\mathcal{E}_V$ & $\mathcal{E}_{\delta}(\vec{x}_0)$\\
\br
FMM Int Src (0.01\%) & 11.5113 & 0.23\% & 0.57\% \\
FMM Int Src (0.1\%) & 8.44722 & 0.39\% & 0.94\% \\
FMM Int Src (1\%) & 5.93731 & 1.91\% & 3.80\% \\
FMM Int Src (10\%) & 3.44240 & 15.11\% & 22.36\% \\
\br
\end{tabular}
\end{indented}
\end{table}

Similarly, we reuse the two examples ($p_{_{O_2}}=10, 160$\,Torr) in Section \ref{DiffP} to study the effect of adaptive strategies for different pressures, and the results are shown in tables \ref{tadaptivep1} and \ref{tadaptivep2}. The performance of the speed-or-accuracy trade-off is similar to the one in different sizes of domain, which validates the robustness of adaptive strategy \eqref{eq:sourcepct} to different pressures as well as different sizes of domain. To ensure the relative error is around $1\%$, $I_{pct}=0.1\%$ is a reasonable parameter. 

\Table{\label{tadaptivep1}Time usage and relative error of FMM Int Src method with different $I_{pct}$ in the adaptive strategy \eqref{eq:sourcepct}, for the case of one Gaussian in (\ref{po2init}) with $p_{_{O_2}}=160$\,Torr. $\vec{x}_0$ is the center of domain $V$ and $\delta = 0.05$\,cm.} 
\br
Method & Time usage (s) &$\mathcal{E}_V$ &$\mathcal{E}_{\delta}(\vec{x}_0)$ \\
\br
FMM Int Src ($0.01\%$) & 10.2820 & 0.16\% & 1.33\% \\
FMM Int Src ($0.1\%$) & 8.38895 & 0.25\% & 1.89\% \\
FMM Int Src ($1\%$) & 4.75543 & 1.50\% & 5.66\% \\
FMM Int Src ($10\%$) & 4.27383 & 13.97\% & 27.30\% \\
\br
\end{tabular}
\end{indented}
\end{table}

\Table{\label{tadaptivep2}Time usage and relative error of FMM Int Src method with different $I_{pct}$ in the adaptive strategy \eqref{eq:sourcepct}, for the case of one Gaussian in (\ref{po2init}) with $p_{_{O_2}}=10$\,Torr. $\vec{x}_0$ is the center of domain $V$ and $\delta = 0.05$\,cm.} 
\br
Method & Time usage (s) &$\mathcal{E}_V$ &$\mathcal{E}_{\delta}(\vec{x}_0)$\\
\br
FMM Int Src ($0.01\%$) & 10.3019 & 0.21\% & 0.49\% \\
FMM Int Src ($0.1\%$) & 8.43274 & 0.35\% & 0.82\% \\
FMM Int Src ($1\%$) & 4.83893 & 1.83\% & 3.54\% \\
FMM Int Src ($10\%$) & 4.17738 & 15.88\% & 22.95\% \\
\br
\end{tabular}
\end{indented}
\end{table}

The selection of adaptive strategy is important. It should be remarked that there are some other adaptive strategies to choose source points besides \eqref{eq:sourcepct}. Nevertheless, as an initial trial to include the adaptive strategy in FMM classical Int method, we think the the discussion of optimal strategy is beyond this work and should be done in the future.

\section{Results and comparison for computing streamer discharges in air}\label{comparison1}
To further illustrate the performance of the adaptive FMM, we study the propagation of streamers in air, where $S_{\rm ph}$ appears as a source term in the fluid model. The governing system of equations is given as (e.g. \cite{Bourdon2010}):
\begin{align}
\left\{ \begin{aligned}
& \frac{\partial n_e}{\partial t} - \nabla \cdot (\mu_e \vec{E} n_e) - \nabla \cdot (D_e \nabla n_e) = S_i - S_{\rm att} + S_{\rm ph}, \\
& \frac{\partial n_p}{\partial t} + \nabla \cdot (\mu_{p} \vec{E} n_p) = S_i + S_{\rm ph}, \\
& \frac{\partial n_n}{\partial t} + \nabla \cdot (\mu_{n} \vec{E} n_n) = S_{\rm att}, \\
& {- \Delta \phi} = \frac{e}{\varepsilon_0} (n_p-n_e - n_n), \qquad \vec{E} = -\nabla \phi,
\end{aligned}
\right.
\label{eM}
\end{align}
where $n_e$, $n_p$ and $n_n$ are the densities of electrons, positive ions and negative ions, respectively;  $\phi$ and $\vec{E}$ denote the electric potential and electric field, respectively; $\mu_e$, $\mu_p$ and $\mu_n$ are the mobility coefficients for electrons, positive ions and negative ions, respectively; $D_e$ is the diffusion coefficient; $e$ and $\varepsilon_0$ are the elementary charge and the vacuum dielectric permittivity, respectively. The three source terms in \eqref{eM} are: the ionization $S_i = \mu_e |\vec{E}| \alpha n_e $, the electron attachment $S_{\rm att} = \mu_e |\vec{E}| \eta n_e$ and the photoionization $S_{\rm ph}$, where $\alpha$ and $\eta$ are ionization and attachment coefficients, respectively. The photoionization rate $S_{\rm ph}$ is given in \eqref{integral}-\eqref{gfun} with $S_i$ in \eqref{eM}. 
In this paper, all the transport coefficients ($\mu_{e,p,n}$ and $D_e$) and reaction rates ($\alpha$ and $\eta$) for the air at atmospheric pressure ($760$\,Torr) are taken from \cite{Bourdon2010, Morrow1997}. The other parameters in \eqref{integral}-\eqref{gfun} follow our previous setting \cite{Lin2020psst}: $p_q =30$\,Torr, $\xi=0.1$, $\omega/ \alpha = 0.6$.

We perform a numerical simulation of one positive streamer interacting with localized plasma region or plasma cloud in three dimensions, as studied in \cite{Babaeva2018, Yuan2020}. This localized region could arise from the preceding streamer channel in a repetitive discharge. 

The main purpose of this section is to compare the adaptive FMM with other methods in the propagation of streamer, and interested readers may refer to \cite{Babaeva2018, Yuan2020} for the effect of different plasma clouds.

The computational domain is set to be a three-dimensional cuboid as $V = [0, 0.5] \times [0, 0.5] \times [0,1.2]$\,cm$^3$, and the streamer discharge between two parallel planar electrodes is simulated. For the Poisson equation, the Dirichlet boundary conditions are applied on the upper face as $\phi |_{z=1.2}=60$\,kV and the bottom face as $\phi|_{z = 0}=0$\,kV; and the homogeneous Neumann boundary conditions are applied on the other four faces. Homogeneous Neumann boundary conditions are adopted on all boundaries for $n_e$ and the inflow boundaries for $n_p$ and $n_n$. 

We follow our previous work \cite{Lin2020psst, tmag2020} to discretize \eqref{eM} by a second-order finite volume method. The stopping criterion for the algebraic elliptic solver is also identical to \cite{Lin2020psst}, where the relative residual less than $10^{-8}$ for the Poisson equation and $10^{-6}$ for the other elliptic equations about photoionization. The second-order explicit method is adopted for the temporal integration, which evaluates both $\phi$ and $S_{\rm ph}$ twice at each time step. The domain $V$ is partitioned uniformly by $512 \times 512 \times 1280$ cells. Time step is chosen as $2.5$\,ps, and the simulation is performed until $3.5$\,ns. For the parallel computing, 32 nodes (640 cores) are taken in the simulation.

All three mentioned adaptive strategies in section \ref{fmm} are adopted for the simulations, using the FMM Int SrcTrg method. Based on the results in section \ref{chapteradaptive}, the first strategy is taken according to \eqref{eq:sourcepct} with $I_{\rm pct} = 0.1\%$. For the second strategy, we choose those $\vec{x}_i$ in \eqref{summationI} which satisfies
\begin{displaymath}
|\vec{E}(\vec{x}_i)| \geq E_{\rm applied} + 0.5\% \left( E_{\max} - E_{\rm applied} \right),
\end{displaymath}
where $E_{\rm applied}$ is the magnitude of the applied electric field and $E_{\max}$ is the maximum value of $|\vec{E}|$. In this section, $E_{\rm applied} = 50$\,kV$\cdot$cm$^{-1}$. This strategy focuses on the region where the electric field is larger than a threshold. This region contains those points $\vec{x}_i$ ahead of streamer, which need photoionization to provide seed electrons during the propagation of a positive streamer. Here we adopt a relatively small parameter (i.e.,0.5\%) to ensure the accuracy, and readers may increase this number to gain more efficiency. Finally, as mentioned in section \ref{fmm}, we adopt the third strategy with a prescribed number $r_s = 10^{-4}$.

Two initial settings are considered in two following sections. The first is the propagation of one positive streamer without plasma cloud, and the other is the interaction of a positive streamer with ion-ion plasma cloud. 

\subsection{Positive streamer without plasma cloud}\label{sec: positivestreamer}
In this example, we compare the different approximations of photoionization in the propagation of one positive streamer. The initial values for $n_e$ and $n_p$ are equal (i.e., $n_e(\vec{x},t=0) = n_p(\vec{x},t=0)$), which are set as a Gaussian
\begin{equation}
10^{14} \exp\left(- \left( (x-0.25)^2 + (y-0.25)^2 + (z-1.2)^2\right) / (0.04)^2 \right) \text{cm}^{-3},
\label{eq:firstGuass}
\end{equation}
and the initial value for $n_n$ is set as zero. This initial setting places a electrically neutral plasma, composed of positive ions and electrons, at the center of anode. 

\begin{figure}[htb!]
	\centering
	\includegraphics[width=0.48\textwidth]{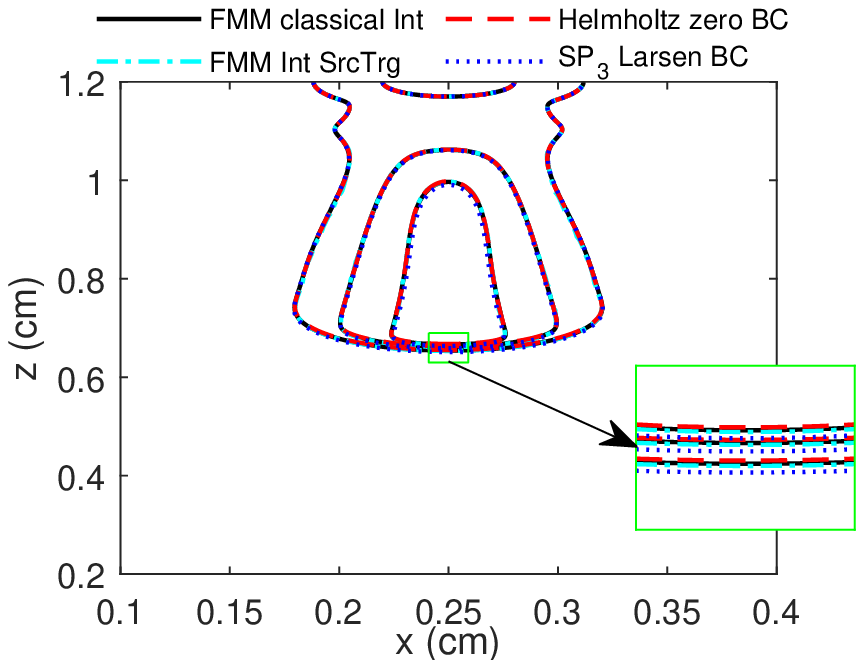}
	\includegraphics[width=0.48\textwidth]{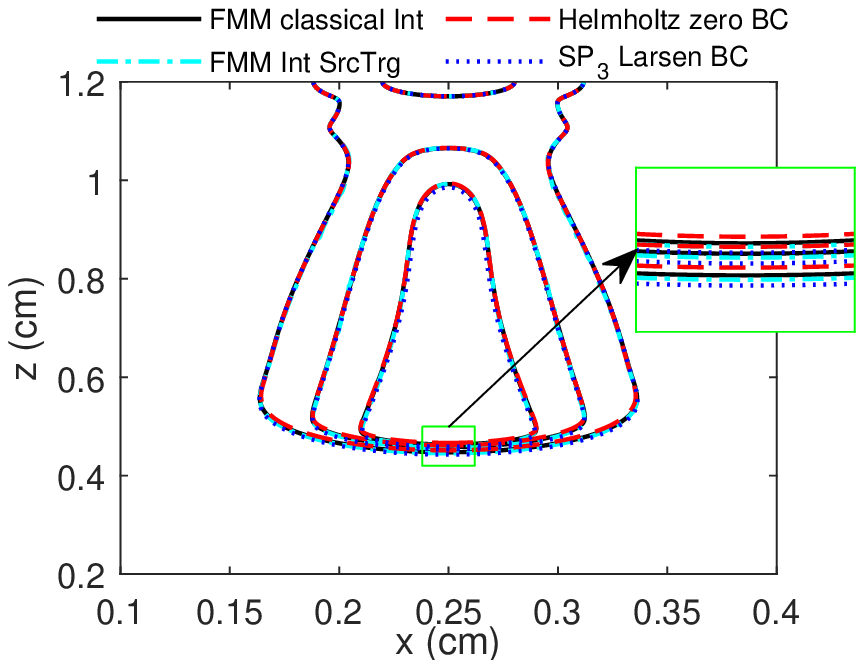}
	\caption{\label{fig:NopSigma}Contours of electron density for positive streamer without plasma cloud. Three contours $n_e=1 \times 10^{13}$, $6 \times 10^{13}$, $1.1 \times 10^{14}$\,cm$^{-3}$ are plotted on plane $y=0.25$\,cm at $3$\,ns (left) and $3.5$\,ns (right).}
\end{figure}

\begin{figure}[htb!]
	\centering
	\includegraphics[width=0.48\textwidth]{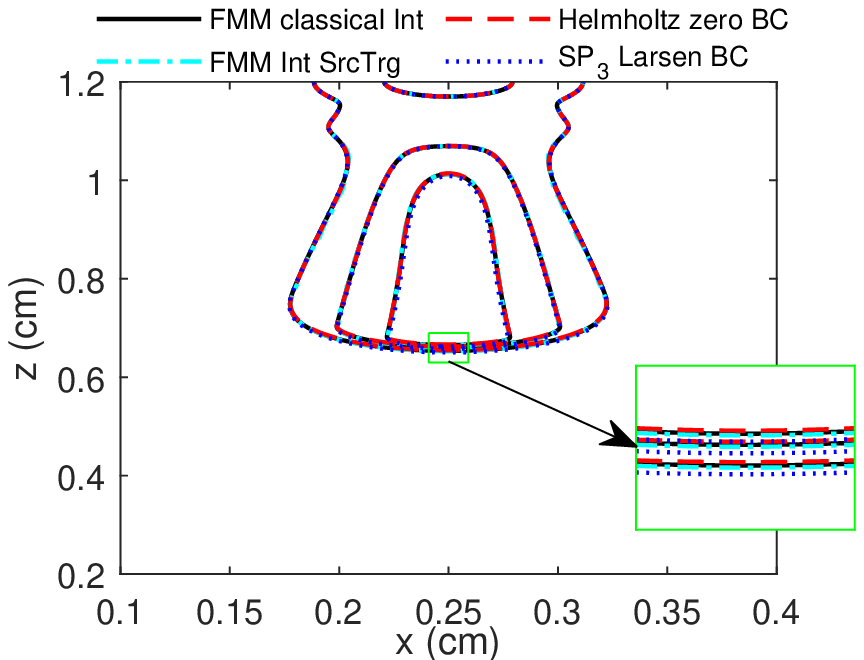}
	\includegraphics[width=0.48\textwidth]{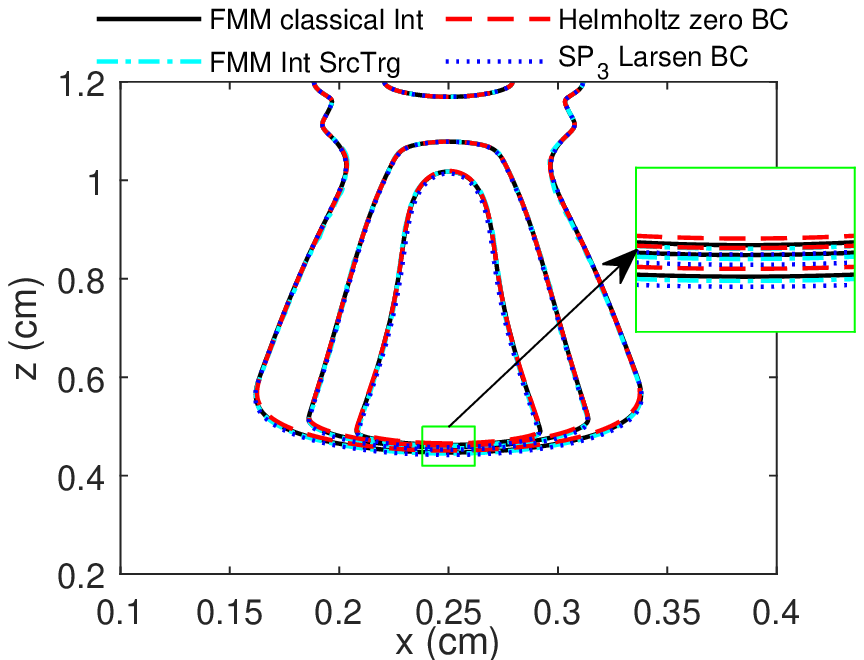}
	\caption{\label{fig:NopP}Contours of the density of positive ions for positive streamer without plasma cloud. Three contours $n_p=1 \times 10^{13}$, $6 \times 10^{13}$, $1.1 \times 10^{14}$\,cm$^{-3}$ are plotted on plane $y=0.25$\,cm at $3$\,ns (left) and $3.5$\,ns (right).}
\end{figure}

\begin{figure}[htb!]
	\centering
	\includegraphics[width=0.48\textwidth]{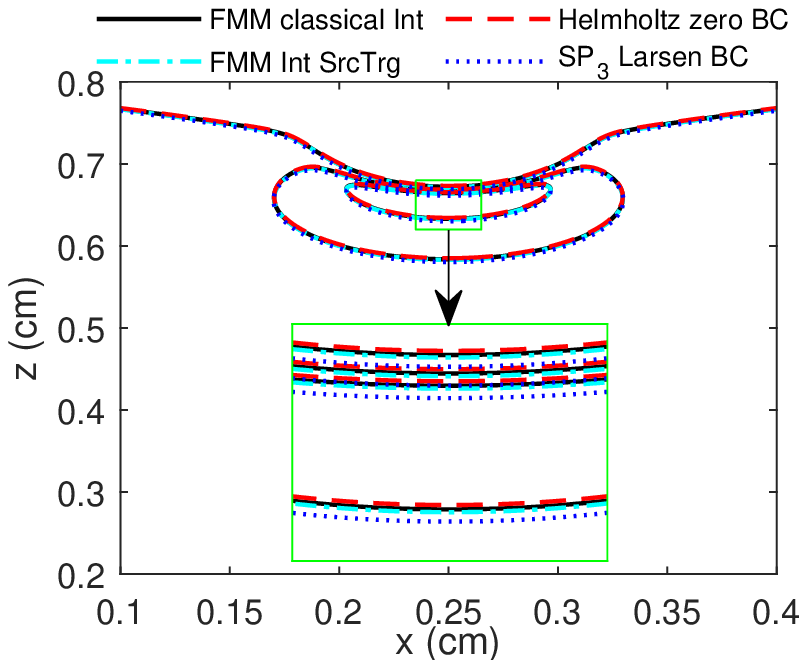}
	\includegraphics[width=0.48\textwidth]{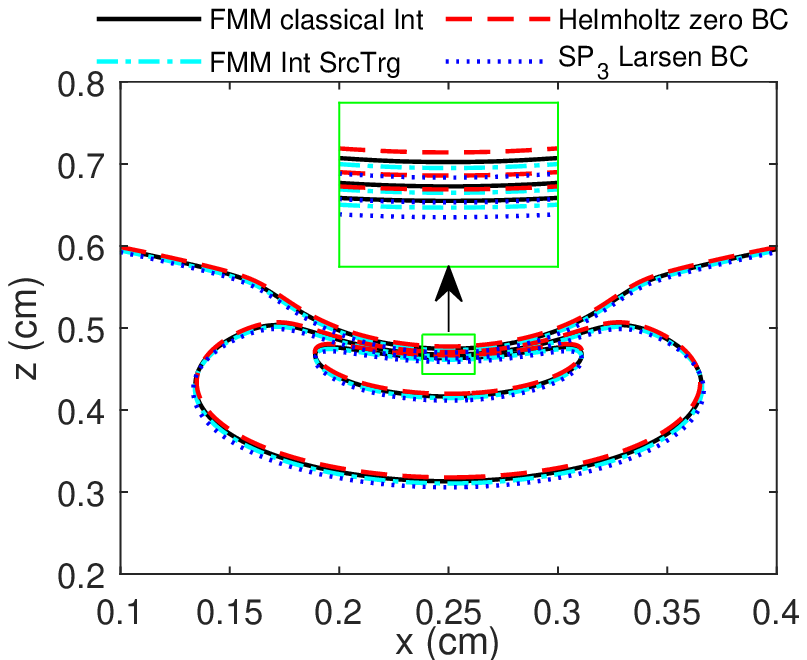}
	\caption{\label{fig:NopEz}Contours of electric field on $z$ direction for positive streamer without plasma cloud. Three contours $E_z=-124.8$, $-83.2$, $-41.6$\,kV$\cdot$cm$^{-1}$ are plotted on plane $y=0.25$\,cm at $3$\,ns (left) and $3.5$\,ns (right).}
\end{figure}

The contours of electron density $n_e$, the density of positive ions $n_p$ and the electric field on $z$ direction $E_z$ are plotted in figures \ref{fig:NopSigma}-\ref{fig:NopEz} at time $t = 3$\,ns and $t = 3.5$\,ns.

An initial observation of figures \ref{fig:NopSigma}-\ref{fig:NopEz} is that the results of all four different methods are in good agreement generally, which is similar to the observation in section 5.1 in \cite{Lin2020psst}. This good agreement can be attributed to the weaker influence of the photoionization compared to the impact ionization. Nevertheless, the differences of four methods are more obvious: 1) near the head of streamer; 2) for longer simulation time. The region near the head of streamer is zoomed in the same figure, where we are easier to distinguish the deviation of the two PDE-based methods to the two FMM methods. Comparing the right sub-figures to the left sub-figures in figures \ref{fig:NopSigma}-\ref{fig:NopEz}, this deviation becomes larger as time increases from $3$\,ns to $3.5$\,ns, which indicates the accurate calculation of the photoionization could be important for a long-time simulation.

Besides observing the deviation of the PDE-based methods to the FMM methods, it is more interesting to focus on the comparison between the FMM classical Int method and its adaptive version, which is the FMM Int SrcTrg method. The contours of the FMM Int SrcTrg method are still greatly close to the ones of the FMM classical Int method even in the zoomed regions and at the longer time. This validates the accuracy of the FMM Int SrcTrg method with the inclusion of mentioned adaptive strategies.

The time cost using four different methods are summarized in table \ref{tab:nocloud}. The FMM Int SrcTrg method is most efficient method among the four methods, which requires least computational cost as shown in table \ref{tab:nocloud}. With the help of adaptive strategies, the FMM Int SrcTrg method reduces one third of the computational times compared with the FMM classical Int method, and is more efficient than the other two PDE-based methods. Therefore, combining with the accurate performance in figures \ref{fig:NopSigma}-\ref{fig:NopEz}, the FMM Int SrcTrg method provides an accurate and efficient way to compute photoionization.

\Table{\label{tab:nocloud}Time usage of simulations using four different methods for the positive streamer without plasma cloud.} 
\br
Method & Time usage (s) \\
\br
FMM classical Int & $1.2088 \times 10^{5}$ \\
FMM Int SrcTrg & $7.8396 \times 10^{4}$ \\
Helmholtz zero BC & $8.7043 \times 10^{4}$ \\
SP$_3$ Larsen BC & $1.5567 \times 10^{5}$ \\
\br
\end{tabular}
\end{indented}
\end{table}

\subsection{Positive streamer with ion-ion plasma cloud}
We would like to study the performance of different photoinization solvers when the positive streamer interacts with a plasma cloud consisting of positive and negative ions. In this example, the positive streamer is initialized as a Gaussian at the center of anode, which is identical to the setting in Section \ref{sec: positivestreamer}. Besides this Gaussian, we place another Gaussian consisting of positive and negative ions at $z=0.6$\,cm, which reads
\begin{equation}
10^{14} \exp\left(- \left( \left( (x-0.25)^2 + (y-0.25)^2 \right) / (0.04)^2 + (z-0.6)^2 / (0.16)^2 \right) \right) \text{cm}^{-3}.
\label{eq:secondGuass}
\end{equation}
Therefore, $n_e(\vec{x},t=0)$ is set as \eqref{eq:firstGuass}; $n_p(\vec{x},t=0)$ is set as a summation of \eqref{eq:firstGuass} and \eqref{eq:secondGuass}; $n_n(\vec{x},t=0)$ is set as \eqref{eq:secondGuass}. 

\begin{figure}[htb!]
	\centering
	\includegraphics[width=0.48\textwidth]{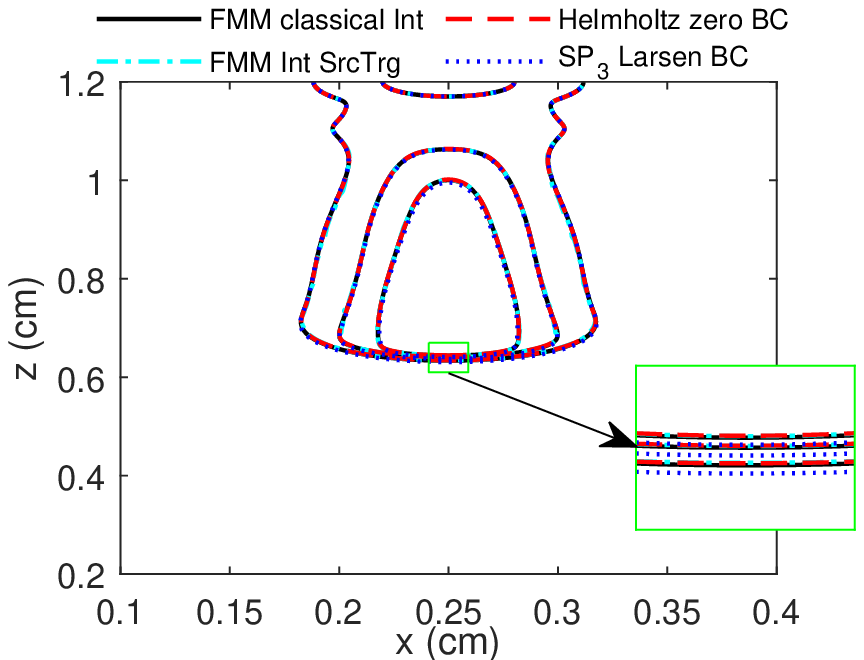}
	\includegraphics[width=0.48\textwidth]{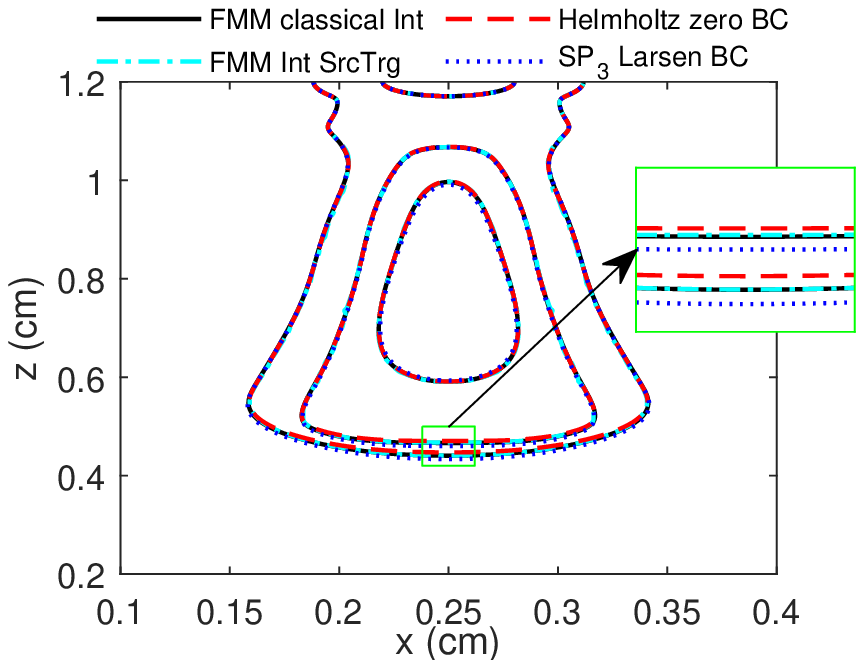}
	\caption{\label{fig:PSigma}Contours of electron density for positive streamer with ion-ion plasma cloud. Three contours $n_e=1 \times 10^{13}$, $6 \times 10^{13}$, $1.1 \times 10^{14}$\,cm$^{-3}$ are plotted on plane $y=0.25$\,cm at $3$\,ns (left) and $3.5$\,ns (right).}
\end{figure}

\begin{figure}[htb!]
	\centering
	\includegraphics[width=0.48\textwidth]{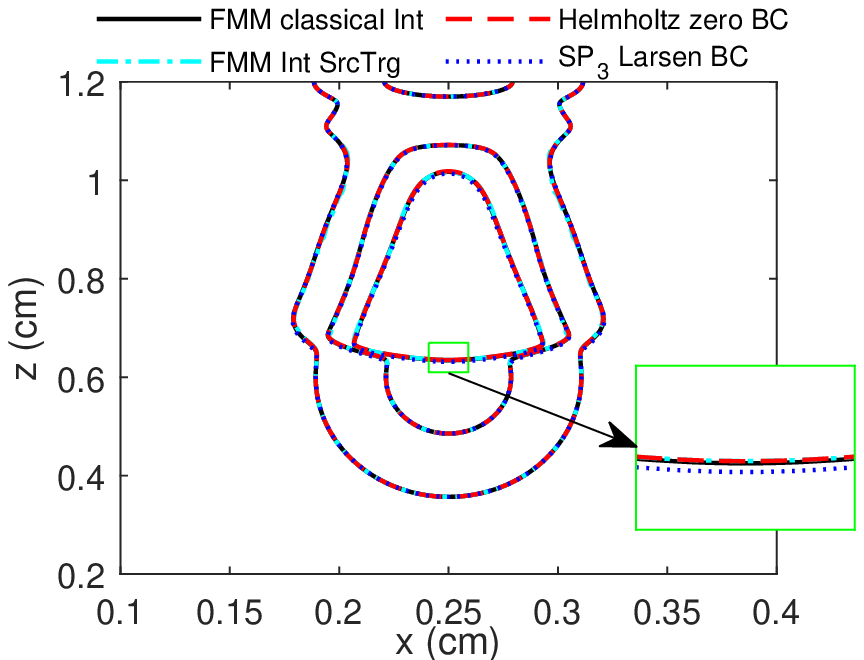}
	\includegraphics[width=0.48\textwidth]{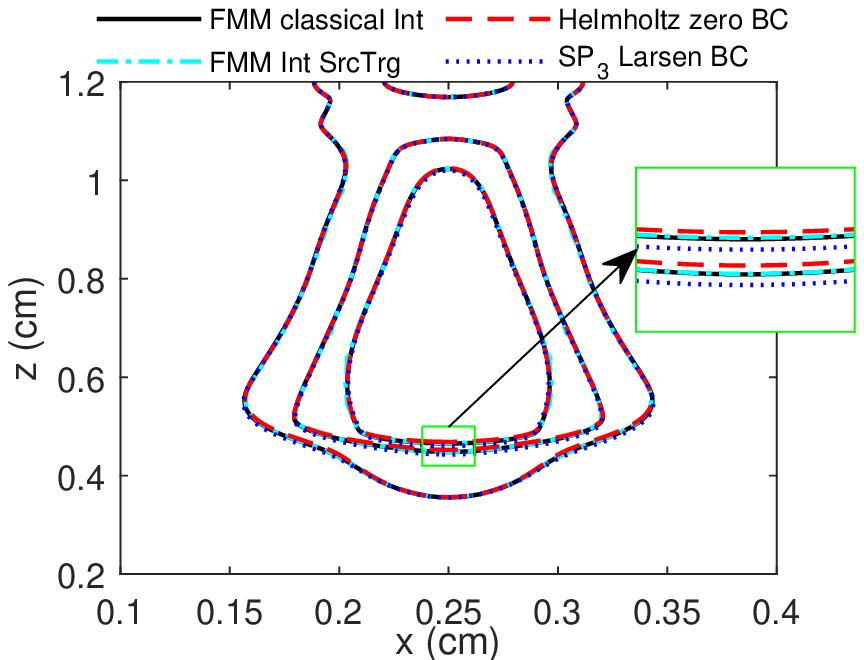}
	\caption{\label{fig:PP}Contours of the density of positive ions for positive streamer with ion-ion plasma cloud. Three contours $n_p=1 \times 10^{13}$, $6 \times 10^{13}$, $1.1 \times 10^{14}$\,cm$^{-3}$ are plotted on plane $y=0.25$\,cm at $3$\,ns (left) and $3.5$\,ns (right).}
\end{figure}

\begin{figure}[htb!]
	\centering
	\includegraphics[width=0.48\textwidth]{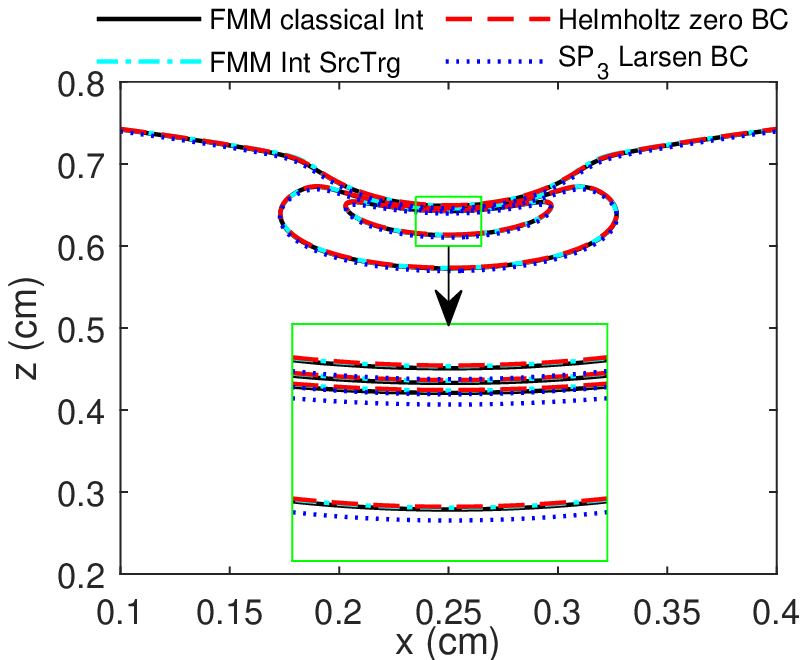}
	\includegraphics[width=0.48\textwidth]{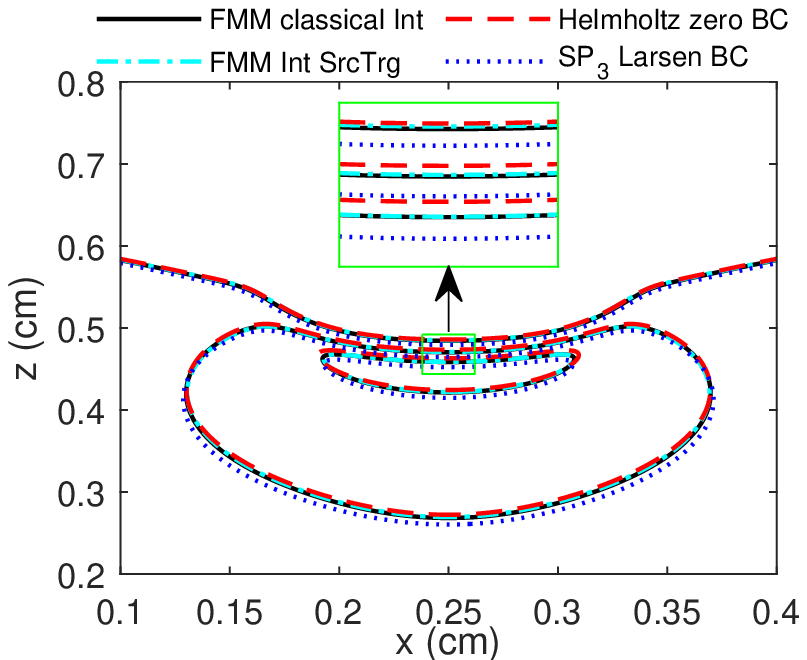}
	\caption{\label{fig:PEz}Contours of electric field on $z$ direction for positive streamer with ion-ion plasma cloud. Three contours $E_z=-124.8$, $-83.2$, $-41.6$\,kV$\cdot$cm$^{-1}$ are plotted on plane $y=0.25$\,cm at $3$\,ns (left) and $3.5$\,ns (right).}
\end{figure}

Similar to the previous section, we plot the contours of electron density, density of positive ions and electric field $E_z$ in figures \ref{fig:PSigma}-\ref{fig:PEz} and summarize the time cost in table \ref{tab:pcloud}.

\Table{\label{tab:pcloud}Time usage of simulations using four different methods for the positive streamer with ion-ion plasma cloud.} 
\br
Method & Time usage (s) \\
\br
FMM classical Int & $1.2013 \times 10^{5}$ \\
FMM Int SrcTrg & $7.3452 \times 10^{4}$ \\
Helmholtz zero BC & $7.5191 \times 10^{4}$ \\
SP$_3$ Larsen BC & $1.6015 \times 10^{5}$ \\
\br
\end{tabular}
\end{indented}
\end{table}

The performance of different methods in figures \ref{fig:PSigma}-\ref{fig:PEz} is similar to the one in figures \ref{fig:NopSigma}-\ref{fig:NopEz}. The difference between the contours of FMM Int SrcTrg method and the contours of FMM classical Int method is almost indistinguishable, while the other two PDE-based methods show some deviation especially in the zoomed region and at larger time. This confirms the accuracy of the FMM Int SrcTrg method when streamer interacts with a plasma cloud.

Besides similar performance of accuracy, the efficiency of different methods in table \ref{tab:pcloud} is also similar to the case of positive streamer in table \ref{tab:nocloud}. The inclusion of adaptive strategies speeds up the calculation using FMM and make its time cost comparable to that of the efficient Helmholtz zero BC method.

Although the comparison of positive streamer with and without plasma cloud is not the main topic in this work, we would like to brief our observation on it. Figures \ref{fig:NopSigma}-\ref{fig:PEz} shows the ion-ion plasma cloud has little effect on the velocity and width of the propagating positive streamer. This result coincides with the numerical experiment in section 2.2 in \cite{Babaeva2018}. The authors attribute this weak effect to the lower mobility of ions and therefore smaller effect on redistributing electric field. Although the difference is small, we can observe that the ion-ion plasma cloud slows down the propagation of positive streamer and spreads the head of streamer. Moreover, the electric field varies slower when the positive streamer interacts with the plasma cloud.

\section{Conclusion}\label{conclusion}
This paper focuses on the efficiency and accuracy to compute the photoionization, and proposes adaptive strategies to accelerate the calculation of the fast multipole method in computing photoionization rate. The adaptive strategies filter the computing points in the fast multipole method according to their intensities of source radiation and magnitude of electric field. 

Quantified error and computation time of the fast multipole method with different strategies are studied in comparison of the direct quadrature of the classical integral, approximations using partial differential equations and the fast multipole method without adaptivity. The results illustrate that the fast multipole method with suitable strategies maintain the high accuracy of the original one. More importantly, these strategies help to reduce the computational cost greatly so that the fast multipole method becomes the most efficient one among the methods in comparison. Furthermore, the adaptive strategies show their robustness with respect to domain sizes and pressures. Therefore, the adaptive strategies make the fast multipole method more attractive in computing the phototionization rate not only for its accuracy but also for its efficiency. 

Future works include finding some better strategies, applying it to other integral models and extending it to a stochastic version.

\section*{Acknowledgments} 
Some computations were done on the Tianhe2-JK cluster at Beijing Computational Science Research Center under the kind support of Dr. Lizhen Chen.

\section*{References}

\end{document}